\documentclass[preprint, aps, prl, showpacs, nofootinbib, unsortedaddress]{revtex4-1}

\usepackage{amsfonts, amsmath, amssymb,latexsym}
\usepackage{graphicx} 
\usepackage{xcolor,microtype}
\usepackage{subfig}

\usepackage{hyperref}
\hypersetup{
    breaklinks,
    colorlinks=true,
    linkcolor=blue,
    citecolor=purple,
  urlcolor=purple,
    pdfstartview={FitH},
    pdfview={FitH 0},
 }

\usepackage{bm}
\usepackage{eucal}      

\usepackage{mathrsfs}
\usepackage{soul}
\usepackage{amsbsy}
\usepackage{psfrag}
\usepackage{mathtools}

\def\beq{\begin{equation}}
\def\eeq{\end{equation}}

\def\beq{\begin{equation}}
\def\eeq{\end{equation}}

\definecolor{olivegreen}{rgb}{0,0.6,0}

\def\spacce#1{\hskip #1pt}
\def\drawline#1#2{\raise 2.5pt\vbox{\hrule width #1pt height #2pt}}
\def\solid{\drawline{24}{.5}\nobreak}

\def\bdash{\hbox{\drawline{5.8}{.5}\spacce{2}}}

\def\dashed{\bdash\bdash\bdash\nobreak}

\def\trian{\raise 1.25pt\hbox{$\scriptstyle\triangle$}\nobreak}

\def\dtrian{\raise 1.25pt\hbox%
{$\scriptscriptstyle\bigtriangledown$}\nobreak}

\def\squar{\raise 1.25pt\hbox{$\scriptstyle\Box$}\nobreak}

\def\diamon{\raise 1.25pt\hbox{$\scriptstyle\diamond$}\nobreak}

\def\beq{\begin{equation}}
\def\eeq{\end{equation}}

\def\citalajim03{Del \'Alamo \& Jim\'enez (2003)}

\begin{document}

\title{Wall turbulence without modal instability of the streaks}

\author{Adri\'an Lozano-Dur\'an}
\affiliation{Center for Turbulence Research, Stanford University, USA}
\email{adrianld@stanford.edu}

\author{Marios-Andreas Nikolaidis}
\affiliation{Department of Physics, National and Kapodistrian University of Athens, Greece}

\author{Navid C.~Constantinou}
\affiliation{Research School of Earth Sciences and ARC Centre of Excellence for Climate Extremes, Australian National University, Australia}

\author{Michael Karp}
\affiliation{Center for Turbulence Research, Stanford University, USA}

\begin{abstract}  
 Despite the nonlinear nature of wall turbulence, there is evidence
 that the mechanism underlying the energy transfer from the mean flow
 to the turbulent fluctuations can be ascribed to linear
 processes. One of the most acclaimed linear instabilities for this
 energy transfer is the modal growth of perturbations with respect to
 the streamwise-averaged flow (or streaks). Here, we devise a
 numerical experiment in which the Navier--Stokes equations are
 sensibly modified to suppress these modal instabilities. Our results
 demonstrate that wall turbulence is sustained with realistic mean and
 fluctuating velocities despite the absence of streak instabilities.
\end{abstract}

\maketitle

Turbulence is a primary example of a highly nonlinear phenomenon.
Nevertheless, there is ample agreement that the energy-injection
mechanisms sustaining wall turbulence can be partially attributed to
linear processes~\citep{Jimenez2013}.  The different scenarios stem
from linear stability theory and constitute the foundations of many
control and modeling strategies~\citep{Kim2006,Schmid2012}.  One of
the most prominent linear mechanisms is the modal instability arising
from mean-flow inflection points between high and low streamwise
velocity regions, usually referred to as `streaks'. Although the modal
instability of the streak plays a central role in several theories of
the self-sustaining turbulence~\citep{Hamilton1995, Waleffe1997,
  Hwang2011}, other linear mechanisms have also been implicated in the
process~\citep{Schoppa2002,Delalamo2006a,Hwang2010b}. Up to date, the
relative importance of linear growth in sustaining turbulence remains
an open question.  Here, we devise a novel numerical experiment of a
turbulent flow over a flat wall in which the Navier--Stokes equations
are minimally altered to suppress the energy transfer from the mean
flow to the fluctuating velocities via modal instabilities.  Our
results show that the flow remains turbulent in the absence of such
instabilities.

\begin{figure*}
 \begin{center}
   \includegraphics[width=0.98\textwidth]{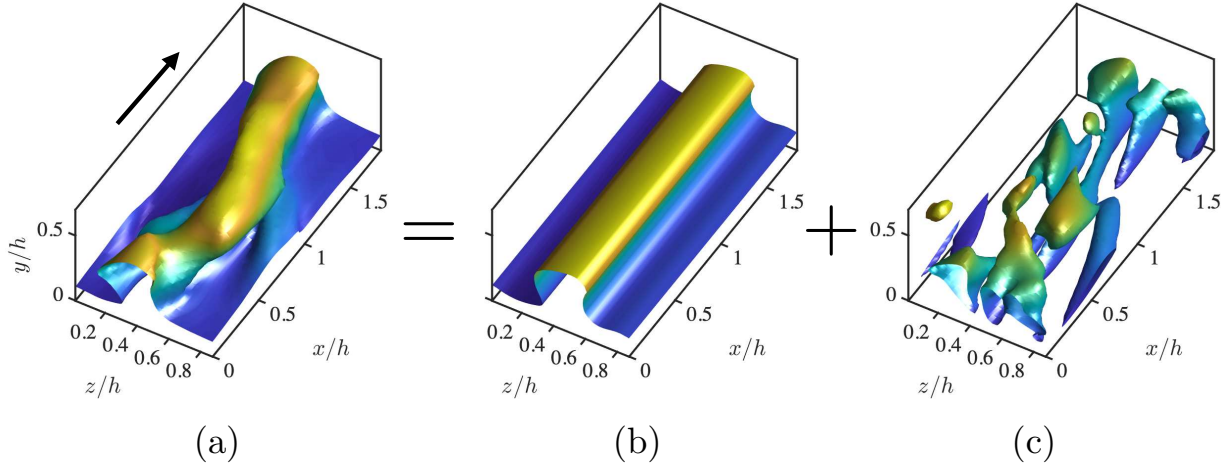}
 \end{center}
\caption{ Decomposition of the instantaneous flow into a streamwise
  mean base state and fluctuations.  Instantaneous isosurface of
  streamwise velocity for (a)~the total flow $u$, (b)~the streak base
  state $U$, and (c)~the absolute value of the fluctuations $|u'|$.
  The values of the isosurfaces are 0.8 (a and b) and 0.1 (c) of the
  maximum streamwise velocity.  Colors represent the distance to the
  wall located at $y=0$. The arrow in panel~(a) indicates the mean
  flow direction.
  \label{fig:snaphots}}
\end{figure*}

Several linear mechanisms have been proposed within the fluid
mechanics community as plausible scenarios to rationalize the transfer
of energy from the large-scale mean flow to the fluctuating
velocities.  Generally, it is agreed that the ubiquitous streamwise
rolls (regions of rotating fluid) and streaks~\citep{Klebanoff1962,
  Kline1967} are involved in a quasi-periodic regeneration
cycle~\citep{Panton2001, Adrian2007, Smits2011, Jimenez2012,
  Jimenez2018} and that their space-time structure plays a crucial
role in sustaining shear-driven turbulence (e.g., Refs.~\cite{Kim1971,
  Jimenez1991, butler1993, Hamilton1995, Waleffe1997, Schoppa2002,
  farrell2012, Jimenez2012, Farrell2016,
  Lozano_brief_2018b}). Accordingly, the flow is often decomposed into
two components: a base state defined by the streamwise-averaged
velocity $U(y,z,t)$ with zero cross-flow (where $y$ and $z$ are the
wall-normal and spanwise directions, respectively), and the
three-dimensional fluctuations (or perturbations) about that base
state. Figure~\ref{fig:snaphots} illustrates this flow decomposition.

Inasmuch as the instantaneous realizations of the streaky flow are
strongly inflectional, the flow $U(y,z,t)$ at a frozen time $t$ is
invariably unstable~\citep{Lozano_brief_2018b}. These inflectional
instabilities are markedly robust and their excitation has been
proposed to be the mechanism that replenishes the perturbation energy
of the turbulent flow~\citep{Hamilton1995, Waleffe1997, Andersson2001,
  Kawahara2003, Hack2014, Hack2018}. Consequently, the modal
instability of the streak is thought to be central to the maintenance
of wall turbulence. The above scenario, although consistent with the
observed turbulence structure~\citep{Jimenez2018}, is rooted in
simplified theoretical arguments.  Whether the flow follows this or
any other combination of mechanisms for maintaining the turbulent
fluctuations remains unclear.
%
%

To investigate the role of modal instabilities, we examine data from
spatially and temporally resolved simulations of an incompressible
turbulent channel flow driven by a constant mean pressure
gradient. Hereafter, the streamwise, wall-normal, and spanwise
directions of the channel are denoted by $x$, $y$, and $z$,
respectively, and the corresponding flow velocity components and
pressure by $u$, $v$, $w$, and $p$. The density of the fluid is $\rho$
and the channel height is $h$. The wall is located at $y=0$, where
no-slip boundary conditions apply, whereas free stress and no
penetration conditions are imposed at $y=h$. The streamwise and
spanwise directions are periodic. The grid resolution of the
simulations in $x$, $y$, and $z$ is $64 \times 90 \times 64$,
respectively, which is fine enough to resolve all the scales of the
fluid motion. Additional details on the numerical setup are offered
in~Ref.~\cite{Lozano_brief_2018b}.

The simulations are characterized by the non-dimensional Reynolds
number, defined as the ratio between the largest and the smallest
length-scales of the flow, $h$ and $\delta_v = \nu/u_\tau$,
respectively, where $\nu$ is the kinematic viscosity of the fluid and
$u_\tau$ is the characteristic velocity based on the friction at the
wall~\citep{Pope2000}.  The Reynolds number selected is
$\mathrm{Re}_\tau = \delta/\delta_v \approx 180$, which provides a
sustained turbulent flow at an affordable computational
cost~\citep{Kim1987}.  The flow is simulated for $100 h/u_\tau$ units
of time, which is orders of magnitude longer than the typical lifetime
of individual energy-containing eddies~\citep{Lozano2014b}. The
streamwise, wall-normal, and spanwise sizes of the computational
domain are $L_x^+ \approx 337$, $L_y^+ \approx 184$, and $L_z^+
\approx 168$, respectively, where the superscript $+$ denotes
quantities normalized by~$\nu$
and~$u_\tau$.~\citeauthor{Jimenez1991}~\cite{Jimenez1991} showed that
turbulence in such domains contains an elemental flow unit comprised
of a single streamwise streak and a pair of staggered quasi-streamwise
vortices, that reproduce the dynamics of the flow in larger
domains. Hence, the current numerical experiment provides a
fundamental testbed for studying the self-sustaining cycle of wall
turbulence.

We focus on the dynamics of the fluctuating velocities
$\boldsymbol{u}' \equiv (u', v', w')$, defined with respect to the
streak base state $U(y,z,t) \equiv L_x^{-1} \int_{0}^{L_x} u(x,y,z,t)
\,\mathrm{d}x$, such that $u' \equiv u- U$, $v' \equiv v$, and $w' \equiv
w$. The fluctuating state vector $\boldsymbol{q}' \equiv
(u',v',w',p'/\rho u_\tau)$ is governed by
\begin{gather} \label{eq:NS}
\mathcal{P}\frac{\partial\boldsymbol{q}'}{\partial t} =
\mathcal{A}(U)\boldsymbol{q}'+ \boldsymbol{N}(\boldsymbol{q}'),
\end{gather}
where $\mathcal{A}$ is the linearized Navier--Stokes operator for the
fluctuating state vector about the instantaneous $U(y,z,t)$ (see
Fig.~\ref{fig:snaphots}b), the operator $\mathcal{P}$ accounts for the
kinematic divergence-free condition $\boldsymbol{\nabla} \cdot
\boldsymbol{u}' = 0$, and $\boldsymbol{N}$ collectively denotes the
nonlinear terms (which are quadratic with respect of fluctuating flow
fields). The corresponding equation of motion for $U(y,z,t)$ is
obtained by averaging the Navier--Stokes equations in the streamwise
direction.

\begin{figure}
 \begin{center}
   \includegraphics[width=0.6\textwidth]{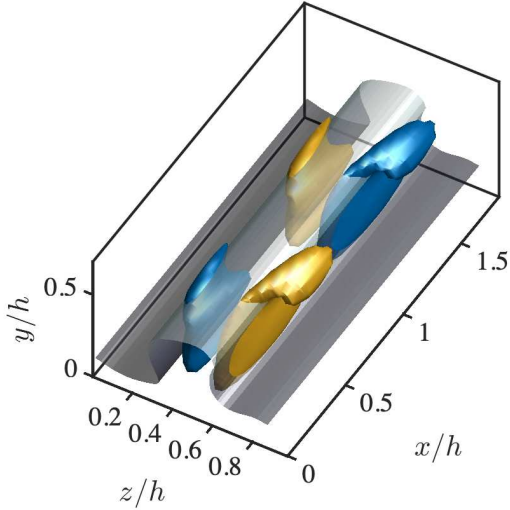}
 \end{center}
\caption{Isosurface of the instantaneous streamwise velocity for the
  eigenmode associated with the most unstable eigenvalue
  $\lambda_\mathrm{max} h/u_\tau \approx 3$ at $t=5.1 h/u_\tau$. The
  values of the isosurface are $-0.5$ (blue) and $0.5$ (yellow) of the
  maximum streamwise velocity. The transparent gray isosurface shows
  the streak at the same instance from
  Fig.~\ref{fig:snaphots}(b). \label{fig:mode1} }
\end{figure}

The modal instabilities of the streaks at a given time are obtained by
eigenanalysis of the matrix representation of the operator
$\mathcal{A}$ about the instantaneous $U$,
\begin{equation}
\mathcal{A}(U) = \mathcal{Q} \Sigma \mathcal{Q}^{-1},
\end{equation}
where $\mathcal{Q}$ consists of the eigenvectors organized in columns
and $\Sigma$ is the diagonal matrix of associated eigenvalues,
$\sigma_i$. The streak is unstable when the growth rate $\lambda_i
\equiv \mathrm{Real}(\sigma_i)$ is positive.  Figure~\ref{fig:mode1}
shows a representative example of the streamwise velocity of an
unstable eigenmode. The predominant eigenmode has the typical sinuous
structure of positive and negative patches of velocity flanking the
velocity streak side by side, which may lead to its subsequent
meandering and breakdown.
%

\begin{figure}
 \begin{center}
 \includegraphics[width=0.95\textwidth]{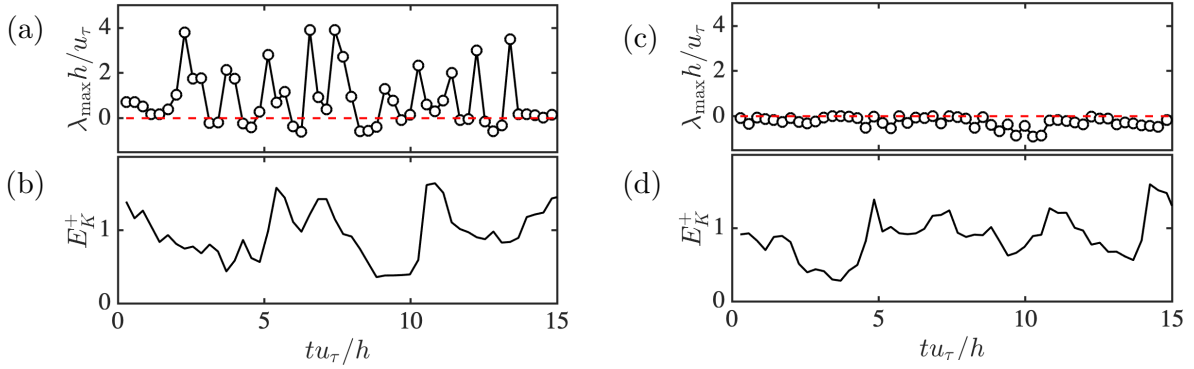}
 \end{center}
\caption{ (a,c) The evolution of the most unstable
  eigenvalue~$\lambda_{\rm max}$ of (a)~$\mathcal{A}$ for the regular
  channel flow and (c)~$\tilde{\mathcal{A}}$ for the channel flow with
  suppressed modal instabilities. (b,d) The evolution of the kinetic
  energy of the perturbations~$E_K=\boldsymbol{u}' \cdot
  \boldsymbol{u}'/2$ averaged over the channel domain for (b)~the
  regular channel flow and (d)~for the channel flow with suppressed
  modal instabilities.
 \label{fig:lambda}}
\end{figure}

We consider two numerical experiments. First, we simulate the
Navier--Stokes equations without any modification, in which the modal
growth of perturbations is naturally allowed. We refer to this case as
the ``regular channel.''  On average, the operator $\mathcal{A}$
contains 2 to 3 unstable eigenmodes at a given
instant. Figure~\ref{fig:lambda}(a) shows the evolution of the maximum
growth rate supported by $\mathcal{A}$ and denoted by
$\lambda_{\mathrm{max}}$. The flow is modally unstable
($\lambda_{\mathrm{max}}>0$) 70\% of the time. The corresponding
kinetic energy of the perturbations averaged over the channel is shown
in Fig.~\ref{fig:lambda}(b).
%
%

For the second numerical experiment, we modify the operator
$\mathcal{A}$ so that all the unstable eigenmodes are rendered neutral
for all times. We refer to this case as the ``channel with suppressed
modal instabilities'' and we inquire whether turbulence is sustained
in this case. The approach is implemented by replacing $\mathcal{A}$
at each time-instance by the modally-stable operator
\begin{equation}\label{Atilde}
\tilde{\mathcal{A}} = \mathcal{Q} \tilde{\Sigma} \mathcal{Q}^{-1},
\end{equation}
where $\tilde{\Sigma}$ is the stabilized version of $\Sigma$ obtained
by setting the real part of all unstable eigenvalues of $\Sigma$ equal
to zero.  We do not modify the equation of motion for $U(y,z,t)$.  The
stable counterpart of $\mathcal{A}$ in Eq.~\eqref{Atilde},
$\tilde{\mathcal{A}}$, represents the smallest intrusion into the
system to achieve modally stable wall turbulence at all times while
leaving other linear mechanisms almost intact.
Figure~\ref{fig:lambda}(c) shows the maximum modal growth rate of
$\tilde{\mathcal{A}}$ at selected times with the instabilities
successfully neutralized. It was verified that turbulence persists
when $\mathcal{A}$ is replaced by $\tilde{\mathcal{A}}$
(Fig.~\ref{fig:lambda}d).

\begin{figure}
 \begin{center}
   \includegraphics[width=0.9\textwidth]{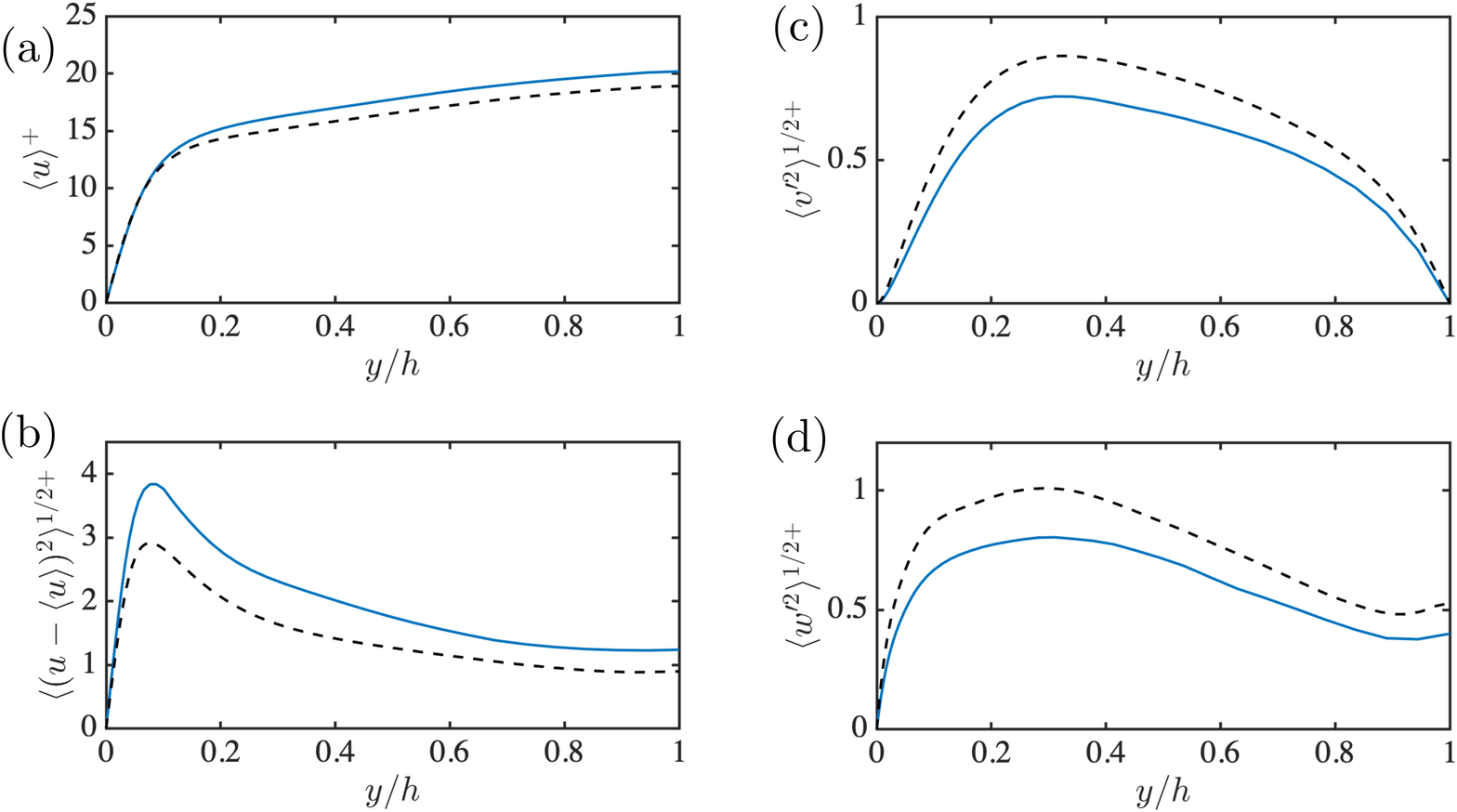}
 \end{center}
\caption{ (a) Streamwise mean velocity profile as a function of the
  wall-normal distance and (b) streamwise, (c) wall-normal, and (d)
  spanwise root-mean-squared fluctuating velocities for the regular
  channel (\dashed) and the channel with suppressed modal
  instabilities (\textcolor{cyan}{\solid}). The Reynolds number of
  both simulations is $\mathrm{Re}_\tau = 186$. Angle brackets
  represent averaging in the homogeneous directions and time.
\label{fig:stats}}
\end{figure}

Our main result is presented in Fig.~\ref{fig:stats}, which compares
the mean velocity profile and turbulence intensities for both the
regular channel and the channel with suppressed modal
instabilities. The statistics are compiled for the statistical steady
state after initial transients.  Notably, the turbulent channel flow
without modal instabilities is capable of sustaining turbulence.  The
difference of roughly 15\%--25\% in the turbulence intensities between
cases indicates that, even if the linear instability of the streak
manifests in the flow, it is not a requisite for maintaining turbulent
fluctuations. The new flow equilibrates at a state with augmented
streamwise fluctuations (Fig.~\ref{fig:stats}b) and depleted cross
flow (Fig.~\ref{fig:stats}c,d). The outcome is consistent with the
occasional inhibition of the streak meandering or breakdown via modal
instability, which enhances the streamwise velocity fluctuations,
whereas wall-normal and spanwise turbulence intensities are diminished
due to a lack of vortices succeeding the collapse of the streak.
%

In summary, we have investigated the linear mechanism of energy
injection from the streamwise-averaged mean flow to the turbulent
fluctuations by modal instabilities.  We have devised a numerical
experiment of a turbulent channel flow in which the linear operator is
altered to render any modal instabilities of the streaks stable, thus
precluding the energy transfer from the mean to the fluctuations via
exponential growth. Our results establish that wall turbulence with
realistic mean velocity and turbulence intensities persists even when
modal instabilities are suppressed. Therefore, we conclude that modal
instabilities of the streaks are not required to attain a
self-sustaining cycle in wall-bounded turbulence. The present outcome
is consequential to comprehend, model, and control the structure of
wall-bounded turbulence by linear methods (e.g.,
Refs.~\cite{Hogberg2003, Delalamo2006a, Hwang2010b, Morra2019}).

Our conclusions refer to the dynamics of wall turbulence in channels
computed using minimal flow units, chosen as simplified
representations of naturally occurring wall turbulence.  The approach
presented in this Letter paves the path for future investigations at
high-Reynolds-numbers turbulence obtained for larger unconstraining
domains, in addition to extensions to different flow configurations in
which the role of modal instabilities remains elusive.

\vspace*{1em}

A.L.-D. acknowledges the support of the NASA Transformative
Aeronautics Concepts Program (Grant~No.~NNX15AU93A) and the Office of
Naval Research (Grant~No.~N00014-16-S-BA10). N.C.C.~was supported by
the Australian Research Council (Grant~No.~CE170100023). This work was
also supported by the Coturb project of the European Research Council
(ERC-2014.AdG-669505) during the 2019 Coturb Turbulence Summer
Workshop at the Universidad Polit\'ecnica de Madrid. We thank Brian
Farrell, Petros Ioannou, Jane Bae, and Javier Jim\'enez for insightful
discussions.

\bibliography{modal_PRL}

\end{document}